\documentclass[aps,prc,groupedaddress,showpacs,manuscript]{revtex4-1}
\usepackage{graphicx}

\usepackage{colordvi}

\begin{document}

\title{Influence of clustering and hadron potentials on the rapidity distribution of protons from the UrQMD model}

\author {Qingfeng Li$\, ^{1}$\footnote{E-mail address: liqf@hutc.zj.cn},
Yongjia Wang$\, ^{1}$,
Xiaobao Wang$\, ^{1}$,
Caiwan Shen$\, ^{1}$, and
Marcus Bleicher$\, ^{2,3}$}

\affiliation{
1) School of Science, Huzhou University, Huzhou 313000, P.R. China \\
2) Frankfurt Institute for Advanced Studies, Ruth-Moufang-Str. 1, 60438 Frankfurt am Main, Germany \\
3) Institut fuer Theoretische Physik, Goethe Universitaet Frankfurt,
Max-von-Laue-Strasse 1, D-60438 Frankfurt am Main, Germany \\
\\
 }
\date{\today}

\begin{abstract}

The Ultra-relativistic Quantum Molecular Dynamics (UrQMD) model supplemented by potentials for both pre-formed hadrons and confined baryons (called UrQMD/M) are used to describe rapidity distributions of both the E895 proton data at AGS energies and the NA49 net proton data at SPS energies. With the help of a coalescence afterburner using only one parameter set of ($R_0$, $P_0$)=(3.8 fm, 0.3 GeV$/$c), both sets of experimental data can be described fairly well except for a small discrepancy seen for the net protons at mid-rapidity from heavy ion collisions (HICs) at high SPS energies. Furthermore, in contrast to the logarithmic dependence with beam energy at SIS energies there are still about 10$\%$ of protons in clusters from central HICs at the beam energy of $80$ GeV$/$nucleon.
\end{abstract}



\maketitle

In order to explore the onset of a possible phase transition from the hadron gas (HG) to the quark-gluon plasma (QGP), the experimental programs have focused on heavy ion collisions (HICs) in the beam energy region below 100 GeV$/$nucleon. These are currently studied experimentally at the BNL Alternating Gradient Synchrotron (AGS), CERN Super Proton Synchrotron (SPS), and with the Beam Energy Scan (BES) program of BNL Relativistic Heavy Ion Collider (RHIC). Lattice quantum chromodynamics (lQCD) calculations \cite{Fodor:2002km} showed that the transition temperature is around 160 MeV (for $\mu_b$ = 0) and the corresponding energy density is around 1 GeV$/$fm$^3$. A multitude of dynamical models have predicted that these temperatures should be reached at the beam energies of 20-30 GeV$/$nucleon. Given the finite baryochemical potential, the heated and compressed nuclear matter created at these energies might pass the phase transition line in the vicinity of a potential critical end point. Therefore, quite a few probes, such as charmonium suppression \cite{Matsui:1986dk}, strangeness enhancement \cite{Soff:1999et}, directed flow \cite{Steinheimer:2014pfa}, elliptic flow (as well as its difference between particles and its anti-partners) \cite{Sorge:1998mk,Torrieri:2007qy,Steinheimer:2012bn,Xu:2013sta,Li:2015aa},  and Hanbury-Brown-Twiss (HBT) two-particle correlation \cite{Adamova:2002ff,Li:2007yd,Li:2010ew,Adamczyk:2014mxp}, have been suggested as signals to detect the possible (phase) transition. And just recently, the baryon stopping itself was proposed to show strong signals of a phase transition as well \cite{Ivanov:2015vna}.

The prerequisite for all studies of hot and dense QCD matter is the deposition of energy and entropy during the initial stage of the nucleus-nucleus reaction. Therefore, the phase space distribution of the protons should be theoretically investigated and well described to benchmark the initial stage. Unfortunately, we observed that the yields of free (net) protons emitted from heavy ion collisions at AGS and SPS energies are not well described in the framework of microscopic transport models such as the Ultra-relativistic Quantum Molecular Dynamics (UrQMD) model \cite{Yuan:2010ad} (using the cascade mode, and called UrQMD/C in this paper), which will be further examined in this paper. The main difference in the calculations and the data is a apparent overestimation of the proton yield in the model.
Here, we explore how a clustering of the baryons into fragments reduces the free proton yield in the model simulation.
This line of argument is suggested by a recent paper of the FOPI group~\cite{Reisdorf:2010aa}, where it was found that for central Au+Au collisions the percentage of free protons is still only about two thirds of the available charge at the beam energy $E_b=1.5$ GeV$/$nucleon. Therefore, the percentage of free protons at higher beam energies such AGS and even SPS deserves attention as well, when one compares model calculations to experimental data.

It is known that, at GSI Schwerionen Synchrotron (SIS) energies, a conventional phase-space coalescence model ~\cite{Kruse:1985pg,Li:2005kqa,Wang:2013wca} can be incorporated with transport models (mainly the QMD-like models) after a proper reaction time $t_{\mathrm{cut}}$ in order to describe multiplicities of clusters and free nucleons. In this afterburner the nucleons with relative momenta $\delta p<P_0$ and relative distances $\delta r<R_0$ will be considered to belong to one cluster. Effects of binding energy, isospin, etc., could be taken into account \cite{Neubert:1999sv,Zhang:2012qm} but are ignored in the current work for simplicity. And, baryons other than nucleons could be treated in a similar  way. Certainly, a powerful afterburner for treating the sequential decays of excited fragments occurring after the dynamic process and for reasonably handling the interface between the dynamic and statistic process is still desirable. In our past calculations, the values for the parameter set ($R_0$, $P_0$) have been chosen in the range of ($R_0=$2.8-3.5 fm, $P_0=$0.25-0.3 GeV$/$c) in order to reproduce experimental data of both multiplicities and collective flows \cite{Li:2002af,Li:2008fn,Russotto:2011hq}. Alternatively, one may apply the Wigner function method explored in \cite{Mattiello:1995xg,Nagle:1996vp,Monreal:1999mv} for the AGS and SPS energy range. While the Wigner function approach seems to be better founded, it has the substantial disadvantage of not conserving baryon number in the projection. Therefore, we employ the traditional approach in this paper. Currently, the coalescence parameters should be varied slightly due to a much higher excitation energy for clusters from heavy ion collisions at AGS and SPS energies. It will be found that only one set of parameters, ($3.8$ fm, 0.3 GeV$/$c), can describe the rapidity distribution of free (net) protons from central Au+Au collisions at AGS and Pb+Pb collisions at SPS energies fairly well, with the help of a mean-field potential version of UrQMD \cite{Li:2007yd,Li:2010ew,Li:2010ie} (called UrQMD/M in this paper). In addition, for each reaction, more than ten thousand events were calculated in the transport program and stopped at $t_{\mathrm{cut}}=50$ fm$/c$.

\begin{figure}[htbp]
\centering
\includegraphics[angle=0,width=0.9\textwidth]{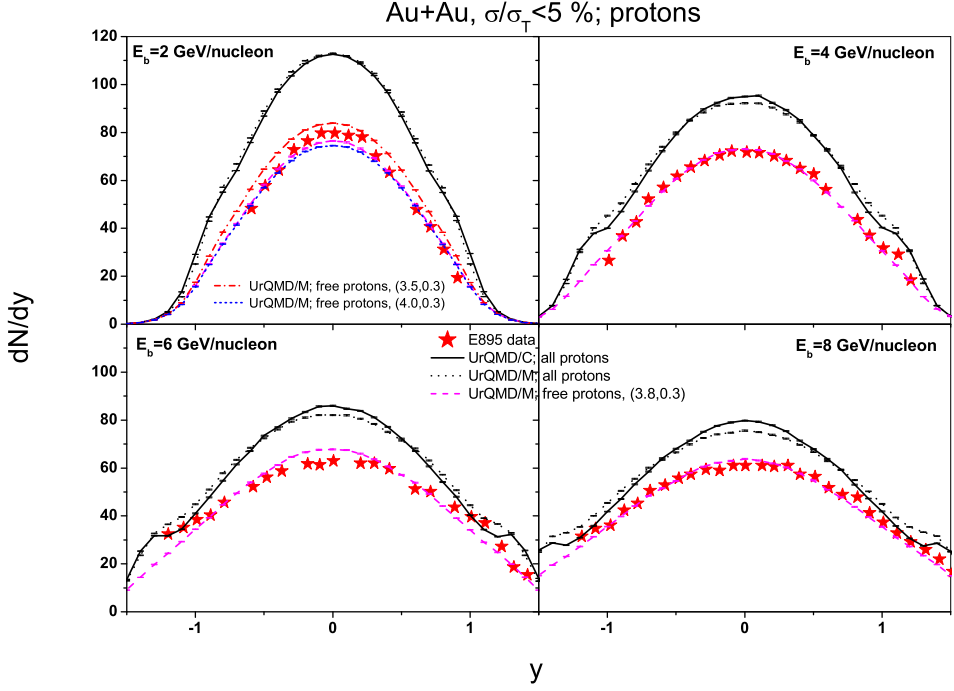}
\caption{\label{fig1} (Color online) Rapidity distribution of protons from central Au+Au reactions at AGS energies 2, 4, 6, and 8 GeV$/$nucleon, respectively. All protons from UrQMD/C and UrQMD/M calculations are shown with solid and dotted lines, respectively. Free protons from UrQMD/M after the coalescence afterburner with the parameter set ($R_0$, $P_0$)=($3.8$ fm, 0.3 GeV$/$c) are shown by dashed lines, while those with parameter sets ($3.5$ fm, 0.3 GeV$/$c) and ($4.0$ fm, 0.3 GeV$/$c) for $E_{b}=$2 GeV$/$nucleon case are shown by dash-dotted and short-dashed lines, respectively. The E895 data are taken from Ref.~\cite{Klay:2001tf}. }
\end{figure}

\begin{figure}[htbp]
\centering
\includegraphics[angle=0,width=0.9\textwidth]{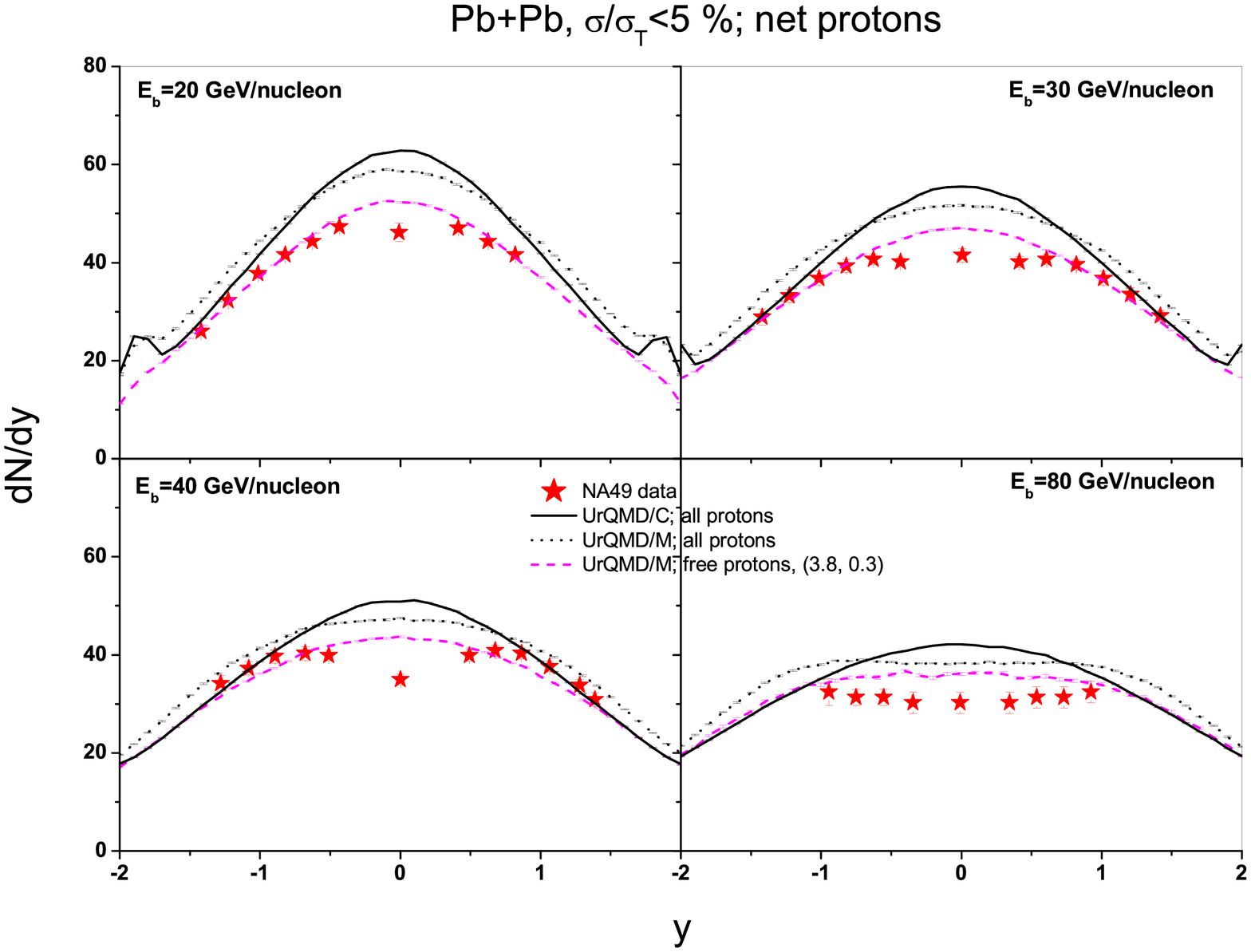}
\caption{\label{fig2} (Color online) Same as Fig.~\ref{fig1} but for net protons from Pb+Pb reactions at SPS energies 20, 30, 40, and 80 GeV$/$nucleon, respectively. The NA49 data are taken from Refs.~\cite{Blume:2007kw,Strobele:2009nq}.
}
\end{figure}

Let us now turn to the rapidity distribution of (free) protons (in Fig.~\ref{fig1}) and (free) net-protons (in Fig.~\ref{fig2}) from central ($\sigma/\sigma_T<5\%$) Au+Au reactions at AGS energies, i.e. $E_{b}=$2, 4, 6, and 8 GeV$/$nucleon, and Pb+Pb reactions at $E_{b}=$20, 30, 40, and 80 GeV$/$nucleon, respectively. UrQMD/C and UrQMD/M calculations are compared to the experimental data by E895 \cite{Klay:2001tf} and NA49 \cite{Blume:2007kw,Strobele:2009nq}. All protons calculated with UrQMD/C and with UrQMD/M are shown by solid and dotted lines, respectively. The free (net) protons from UrQMD/M after the coalescence afterburner with the parameter set ($R_0$, $P_0$)=($3.8$ fm, 0.3 GeV$/$c) are shown with dashed lines. One clearly observes that the mean-field potential modifications for both ``pre-formed'' hadrons and formed baryons in UrQMD/M broadens the rapidity distributions, this in turn leads to a reduction of the proton yield at mid-rapidity. Especially for heavy ion collisions at higher beam energies this effect is more pronounced, as was shown in details in previous calculations \cite{Li:2007yd,Yuan:2010ad}. The additional pressure during the early compression stage leads to a reduced number of subsequent collisions at the later expansion stage and earlier freeze-outs. Next we add the coalescence afterburner to obtain the number of free protons. For comparison, three $R_0$ values 3.5, 3.8, and 4.0 fm are used for varying the yield of free protons and results are only shown for the case with $E_{b}=$2 GeV$/$nucleon. It is clear that the consideration of the afterburner and the increase of $R_0$ value reduce the yields of (net) protons (especially at the low AGS energies) in the whole rapidity region. And, calculations with potentials and the parameter set ($3.8$ fm, 0.3 GeV$/$c) in the coalescence afterburner allows to describe both, the E895 and the NA49 data fairly well. A small discrepancy between UrQMD/M calculations with afterburner and the experimental data at the top SPS energies leaves space for a more systematic description of the dynamical evolution of the new phase created at the early stage, such as the stiffness of EoS \cite{Li:2007yd,Li:2008qm,Xu:2013sta}, effective string tension \cite{Soff:1999et}, and modifications of cross sections \cite{AbdelWaged:2004wk,Xu:2013sta,Steinheimer:2015sha,Li:2015aa}.

\begin{figure}[htbp]
\centering
\includegraphics[angle=0,width=0.9\textwidth]{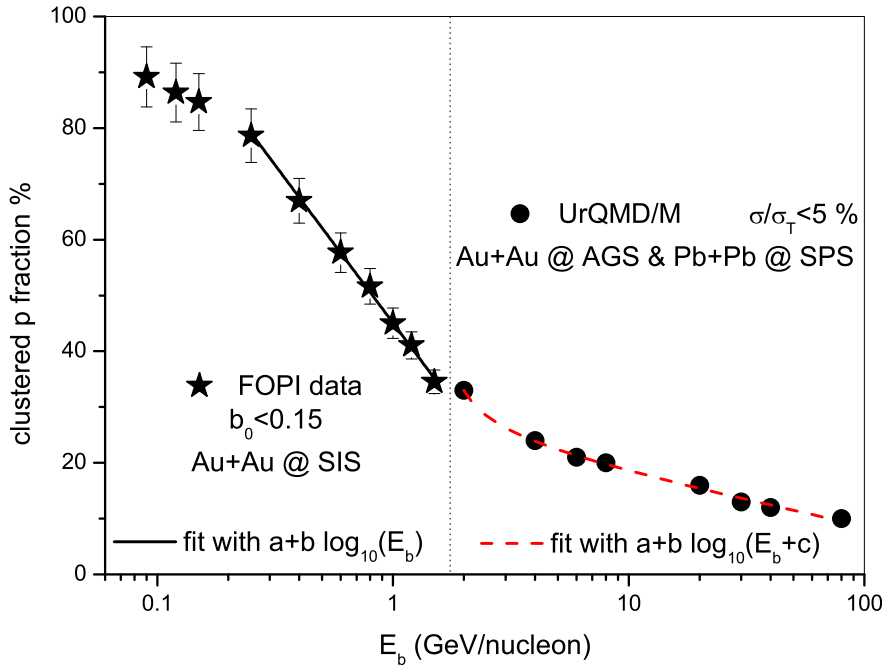}
\caption{\label{fig3} (Color online) Excitation function of the clustered proton fraction (in $\%$). At SIS energies, the FOPI data (stars) are fitted with a function $a+b$log$_{10}$ $(E_{\mathrm b})$ (solid line). At AGS and SPS energies, results from UrQMD/M calculations (circles) are fitted with $a+b$log$_{10}$ $(E_{\mathrm b}+c)$ (dashed line).
}
\end{figure}

As noticed in Ref.~\cite{Reisdorf:2010aa} that, if the proton fraction in clusters to all protons produced from Au+Au collisions at SIS energies and at reduced impact parameters ($b_0<0.15$, where $b_0$ is defined by $b/b_{\mathrm max}$ and $b_{\mathrm max}$ is the sum of both projectile and target sizes) is plotted as a function of beam energy, and the abscissa is set to be logarithmic, it is found that the excitation function shows a nicely linear dependence in the energy range from 0.2 to 1.5 GeV$/$nucleon, which is also shown in the left side of Fig.~\ref{fig3} with scattered star symbols. However, if we extrapolate the fitted line (solid) to higher energies, it would be found that there is no clusters any more at the beam energy around 6.5 GeV$/$nucleon, which is obviously not supported by our UrQMD/M calculations (shown in the right side of Fig.~\ref{fig3} with solid circle symbols) as well as existing experimental results. Furthermore, the logarithmic dependence with beam energy is destroyed with a non-zero value of the extra parameter $c$ in the fitting function (shown by the dashed line). It is interesting to see that the clustered proton fraction from UrQMD/M calculations for central Au+Au reactions at $E_b=2$ GeV$/$nucleon matches that from the FOPI data at 1.5 GeV$/$nucleon. And, on the other side, at $E_b=80$ GeV$/$nucleon the proton percentage in clusters keeps still on the order of 10. It is also found that the free proton fraction at $20-30$ GeV$/$nucleon (about 85$\%$) has reached the percentage of the averaged number of interacting (wounded) nucleons to the total, which is calculated from the Glauber approach \cite{Strobele:2009nq}.

To summarize, we have employed the UrQMD model supplemented by potentials and called UrQMD/M. In UrQMD/M mean-field potentials for both pre-formed hadrons and confined baryons are considered. For the present investigation a coalescence afterburner with parameter set to ($R_0$,$P_0$)=(3.8 fm, 0.3 GeV$/$c) is used. We found that the E895 proton data at AGS energies and the NA49 net proton data at SPS energies can be described reasonably well, if the comparison is performed for the free protons. The calculated excitation function of the proton fraction existing in clusters deviates from a pure logarithmic function as seen by the FOPI collaboration at SIS energies. In contrast to the low energy extrapolation there are still about 10$\%$ of protons in clusters from central HICs at high SPS energies.

\begin{acknowledgements}
We thank Prof. Fuqiang Wang for fruitful discussions and acknowledge support by the computing server C3S2 in Huzhou
University. The work is supported in part by the National
Natural Science Foundation of China (Nos. 11375062, 11275068), the project sponsored by SRF for ROCS, SEM, and the Doctoral Scientific Research Foundation (No. 11447109). M.B. was supported through the Hessian LOEWE-Center HIC for FAIR.
\end{acknowledgements}

\end{document}